\documentclass{elsart}
\usepackage{ifpdf}
\usepackage{graphicx,amssymb,lineno}
\usepackage[square, comma, sort&compress]{natbib}
\usepackage{amsmath}
\usepackage{epsfig}

\begin{document}

\begin{frontmatter}
\title{Precision measurement of the electric quadrupole moment of $^{31}$Al and determination of the effective proton charge in the sd-shell}

\author[a]{M. De Rydt},
\ead{marieke.derydt@fys.kuleuven.be}
\author[a]{G. Neyens},
\author[b]{K. Asahi},
\author[c]{D.L. Balabanski},
\author[d]{J.M. Daugas},
\author[a]{M. Depuydt},
\author[d]{L. Gaudefroy},
\author[e]{S. Gr\'evy},
\author[b]{Y. Hasama},
\author[f]{Y. Ichikawa},
\author[d]{P. Morel},
\author[f]{T. Nagatomo},
\author[i]{T. Otsuka},
\author[g]{L. Perrot},
\author[h]{K. Shimada},
\author[e]{C. St\"odel},
\author[e]{J.C. Thomas},
\author[f]{H. Ueno},
\author[i]{Y. Utsuno},
\author[a]{W. Vanderheijden},
\author[a]{N. Vermeulen},
\author[a]{P. Vingerhoets},
\author[f]{A. Yoshimi}

\vspace{0.2cm}

\address[a]{Instituut voor Kern- en Stralingsfysica, K.U.Leuven, Celestijnenlaan 200D,
B-3001 Leuven, Belgium}
\address[b]{Department of Physics, Tokyo Institute of Technology, 2-12-1 Oh-okayama, Meguro-ku, Tokyo 152-8551, Japan}
\address[c]{Institute for Nuclear Research and Nuclear Energy, Bulgarian Academy of Sciences, BG-1784 Sofia, Bulgaria}
\address[d]{CEA, DAM, DIF, F-91297 Arpajon Cedex, France}
\address[e]{GANIL, DEA/DSM-CNRS/IN2P3, BP 55027, F-14076 Caen Cedex 5, France}
\address[f]{RIKEN Nishina Center, 2-1 Hirosawa, Wako, Saitama 351-0198, Japan}
\address[g]{IPN, 15 Rue G. Clemenceau, F-91406 Orsay, France}
\address[h]{CYRIC, Tohoku University, 6-3 Aoba, Aramaki, Aoba-ku, Sendai, Miyagi 6980-8578, Japan}
\address[i]{Japan Atomic Energy Agency, Tokai, Ibaraki 319-1195, Japan}

\begin{abstract}
The electric quadrupole coupling constant of the $^{31}$Al ground
state is measured to be $\nu_{Q} =
\left|\frac{eQV_{zz}}{h}\right| = 2196(21)\mbox{kHz}$ using two
different $\beta$-NMR (Nuclear Magnetic Resonance) techniques.
For the first time, a direct comparison is made between the continuous rf technique and the adiabatic fast passage method.
The obtained coupling constants of both methods are in excellent agreement with each other and a precise value for the quadrupole moment of
$^{31}$Al has been deduced: $\left|Q({^{31}\mbox{Al})}\right|$ = 134.0(16) mb.
Comparison of this value with large-scale shell-model calculations
in the sd and sdpf valence spaces suggests that the $^{31}$Al ground state is
dominated by normal sd-shell configurations with a possible small
contribution of intruder states. The obtained value for $\left|Q(^{31}\mbox{Al})\right|$ and a compilation of measured quadrupole moments of odd-Z even-N isotopes in comparison with shell-model calculations shows that the proton effective charge $e_p=1.1 e$ provides a much better description of the nuclear properties in the sd-shell than the adopted value $e_p=1.3 e$.
\end{abstract}

\end{frontmatter}

\section{Introduction}
\label{intro}
The electric quadrupole moment \textit{Q} provides a direct measure of the deviation of the nuclear charge distribution from a sphere. As the quadrupole moment is a sensitive probe to indicate changes in the shell structure, its measurement is an important tool in studying nuclei in and near the island of inversion. This region, comprising the Ne (\textit{Z}=10), Na (\textit{Z}=11) and Mg (\textit{Z}=12) isotopes with neutron number around \textit{N}=20, is characterized by the dominance of neutron excitations from the sd to the pf-orbits (known as intruder configurations) in the ground state.\\
The first experimental evidence for this effect was found in the measured nuclear masses and two-neutron separation energies of the $^{26-32}$Na isotopes \cite{Thi75}. In the same year, Campi et al. made the first theoretical analysis of the observed experimental anomalies, performing constrained Hartree Fock calculations \cite{Cam75}. Later, $\beta$-decay studies revealed a low lying 2$^{+}$ state in $^{32}$Mg \cite{Det79,Gui84}, pointing to a deformed ground state in that nucleus. Further evidence for the existence of the island of inversion was provided by the measurement of B(E2)-values and nuclear moments of the neutron-rich Ne, Na and Mg isotopes (e.g. \cite{Pri99,Iwa05, Kei98,Kei00,Tri05, Mot95,Ney05,Sor08}).\\
As new experimental results became available, shell-model calculations grew more important. Starting from the early calculations performed by Watt et al. \cite{Wat81} and Poves and Retamosa \cite{Pov87}, new residual interactions in larger model spaces were developed \cite{Cau98,Uts99} taking into account particle-hole excitations of the neutrons across the reduced \textit{N}=20 shell gap. For nuclei inside the island of inversion, theoretical predictions reproduce the observed nuclear properties quite well. For the neutron-rich $^{31-34}$Al isotopes (\textit{Z}=13) located at the border of the island of inversion, the interpretation of experimental results by means of theoretical models contributes to a better understanding of how the transition to the region of deformed nuclei occurs. A comparison between the measured nuclear magnetic moments of $^{31-34}$Al \cite{Bor02,Uen05,Him06,Him08} and shell model calculations shows that $^{31}$Al and $^{32}$Al can be described as normal sd-shell nuclei \cite{Bor02,Him06}. For $^{33}$Al and $^{34}$Al, intruder configurations play a role in the ground state \cite{Him06,Him08} and in the low-energy level structure \cite{Mit02,Pri01}. This indicates a gradual transition from the Si isotopes, being normal in their ground states, to the deformed Mg isotopes inside the island of inversion.\\
As the nuclear quadrupole moment is even more sensitive to neutron excitations across \textit{N}=20 than the magnetic moment, a measurement of \textit{Q} can provide crucial and decisive information about the mixing between sd and pf configurations in the ground states of the neutron-rich Al isotopes. So far, the \textit{Q}-moments of $^{31-32}$Al have been studied with a limited accuracy in Refs. \cite{Kam07, Nag08}. This letter
reports on the precision measurement of the $^{31}$Al ground-state quadrupole moment, which is a crucial step in describing the systematics of the odd-mass Al isotopes
towards the semi-magic nucleus $^{33}$Al.\\
The quadrupole moment of $^{31}$Al was measured using two different $\beta$-nuclear magnetic resonance techniques, based on a pulsed and a continuous radiofrequent field respectively. This provides a unique opportunity to directly compare both methods and results. The two techniques can be used to determine magnetic moments (then we call it a $\beta$-NMR method) or quadrupole moments (then the term $\beta$-NQR or $\beta$-nuclear quadrupole resonance is adopted).

\section{Experiment}
\label{exp} 
\begin{figure}[htb]
\centering
\includegraphics[scale = 0.31]{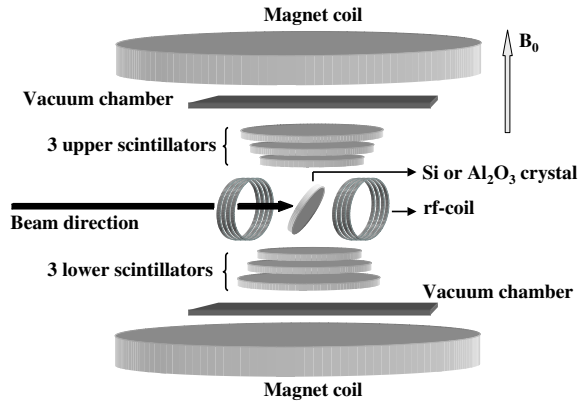}
\caption{A schematic overview of the $\beta$-NMR/NQR set-up.}\label{set_up}
\end{figure}

\begin{figure}[htb]
\centering
\includegraphics[scale = 0.35]{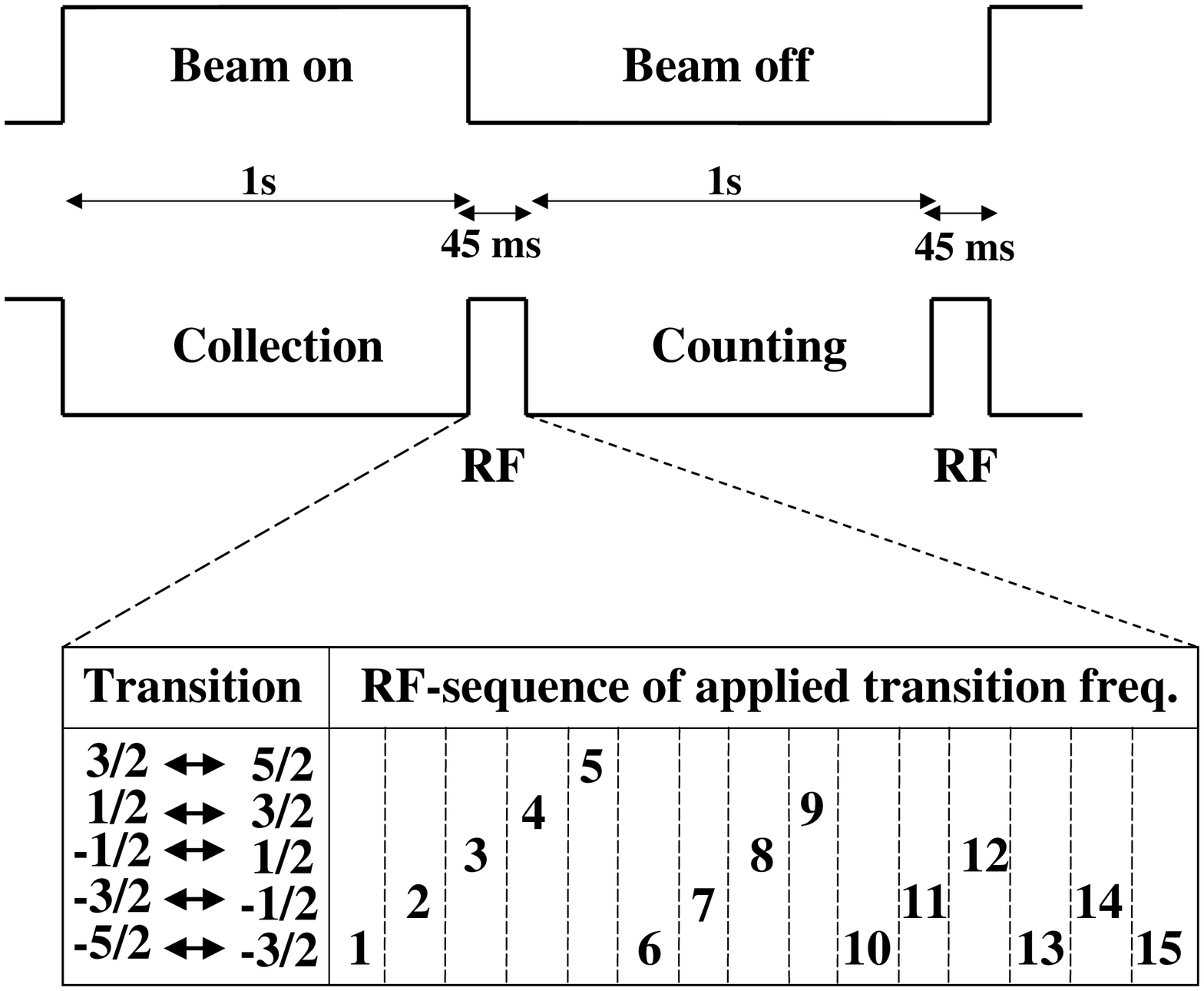}
\caption{Structure of one AFP-NQR cycle. A zoom is made of one rf-on
period. It shows the order in which all five transition frequencies in $^{31}$Al are applied.}
\label{time_scheme}
\end{figure}

Neutron-rich $^{31}$Al nuclei were
produced in a projectile fragmentation reaction, induced by a
$^{36}$S$^{16+}$ primary beam (77 MeV/u) impinging on a 1212
$\mu$m thick $^9$Be target. Selection of the secondary beam was
done with the high-resolution fragment separator LISE at GANIL
\cite{Ann87,Ann92}. Si-diodes have been used for the beam
identification by standard energy loss versus time of flight
measurements. Spin polarization of the fragments \cite{Asa90} was obtained by putting an angle of 2(1)$^\circ$ on the primary beam with respect to the entrance of the spectrometer. To get the highest polarization, a selection in the right wing of the longitudinal momentum distribution was made, giving a polarized $^{31}$Al production yield of about 10$^4$ pps for a primary beam intensity of 1 $\mu$A. A purity of 90\% was achieved using a 1054 $\mu$m thick $^9$Be wedge-degrader in the intermediate focal plane.\\
At the end of the LISE fragment separator, a $\beta$-nuclear
magnetic/quadrupole resonance ($\beta$-NMR/NQR) set-up is
installed. The $^{31}$Al-nuclei are implanted in a cubic
Si-crystal for $\beta$-NMR or in an $\alpha$-Al$_2$O$_3$ (corundum) crystal for
$\beta$-NQR. No crystal cooling is needed since the spin-lattice relaxation time $T_1$ of $^{31}$Al (\textit{t}$_{1/2}$ = 644(25) ms) in both crystals is sufficiently long to preserve the nuclear polarization for a few half lives. $T_1$ of $^{31}$Al in an $\alpha$-Al$_2$O$_3$ crystal is estimated to be 4.8 s, using the $T_1T^2Q^2$ = const law \cite{Mie60} and the measured relaxation time of $^{27}$Al \cite{Ken94}.\\
The implanted nuclei are exposed
to an external magnetic field $B_0$ which
induces a Zeemann splitting of the nuclear \textit{m}-states. When a
non-cubic crystal (e.g. Al$_2$O$_3$) is used as a stopper material,
the quadrupole interaction with the electric field gradient causes
an additional shift of the magnetic substates which results in a
non-equidistant level spacing. Perpendicular to the external
magnetic field, a radiofrequent (rf) field is applied. This rf-field is
generated inside the coil mounted around the crystal, by a
function generator and rf-amplifier. The asymmetry in the
$\beta$-decay of the polarized nuclei is observed in two sets of
three thin plastic scintillators, one set is situated above the
crystal, the other below. Scattering and noise events
are reduced in this configuration by requiring a coincidence
between the three detectors in each set. A schematic drawing of
the $\beta$-NMR/NQR set-up is given in Fig. \ref{set_up}.
\\The transition frequency between the magnetic sublevels \textit{m} and \textit{m}+1
under the influence of a Zeemann and an axially symmetric quadrupole interaction can be calculated using perturbation theory, provided that the Larmor frequency $\nu_L$ is much larger than $\frac{\nu_Q}{4I(2I-1)}$. Up to the second order, this series expansion
is given by \cite{Abr61}:
\begin{equation}
\label{transitionFreq}
\nu_m - \nu_{m+1} = \nu_L - \frac{3\nu_Q}{8I(2I-1)}(2m+1)\left(3\cos^2\theta-1\right)\\
\end{equation}
\begin{equation*}
-\; \frac{\nu_Q^2}{32 \nu_L}\; \frac{9}{4I^2(2I-1)^2}\; \left[6m(m+1)-2I(I+1)+3\right]\; \sin^4\theta
\end{equation*}
\textit{I} is the nuclear ground-state spin and $\theta$ is the angle
between the electric field gradient along the \textit{c}-axis of the crystal and the external magnetic field
$B_0$. The Larmor frequency $\nu_{L}$ equals
$\frac{g\mu_{N}B_0}{h}$ with $g$ the nuclear g-factor. The
quadrupole coupling constant $\nu_{Q}$ is defined as
$\frac{eQV_{zz}}{h}$. In this expression, $Q$ is the spectroscopic quadrupole moment (in mb) and $V_{zz}$
the axially symmetric electric field gradient.
\\For $^{31}$Al (\textit{I}=5/2$^+$), five different transition frequencies exist. In table \ref{table:NQR_freq}, these five frequencies are given in case the symmetry axis of the electric field gradient is parallel ($\theta$=0$^\circ$) or perpendicular ($\theta$=90$^\circ$) to the direction of \textit{B}$_0$. For the 90$^{\circ}$ orientation, the quadrupole splitting is about two times smaller than for the 0$^{\circ}$ direction which implies that a double quadrupole moment window can be scanned with the same frequency settings.
\begin{table}[htb]
\caption{Overview of the transition frequencies between the m-states in $^{31}$Al.}
\begin{tabular}{|c|c|c|}
\hline
\;Transition\; & $\theta=0^{\circ}$ & $\theta=90^{\circ}$ \\
\hline
$-5/2 \leftrightarrow -3/2$\; & \;$\nu_{1} = \nu_L + \frac{3}{10}\nu_Q$ \;& \;$\nu_{1} = \nu_L - \frac{3}{20}\nu_Q + \frac{9}{1600}\left(\frac{\nu_Q^2}{\nu_L}\right)$\\

$-3/2 \leftrightarrow -1/2$\; & $\nu_{2} = \nu_L + \frac{3}{20}\nu_Q$ & \;$\nu_{2} = \nu_L - \frac{3}{40}\nu_Q + \frac{9}{1280}\left(\frac{\nu_Q^2}{\nu_L}\right)$\\

$-1/2 \leftrightarrow 1/2$ & $\nu_{3} = \nu_L$  & $\nu_{3} = \nu_L + \frac{9}{800}\left(\frac{\nu_Q^2}{\nu_L}\right)$\\

$1/2 \leftrightarrow 3/2$ & $\nu_{4} = \nu_L - \frac{3}{20}\nu_Q$ & \;$\nu_{4} = \nu_L + \frac{3}{40}\nu_Q - \frac{9}{1280}\left(\frac{\nu_Q^2}{\nu_L}\right)$\\

$3/2 \leftrightarrow 5/2$ & $\nu_{5} = \nu_L - \frac{3}{10}\nu_Q$ & \;$\nu_{5} = \nu_L + \frac{3}{20}\nu_Q - \frac{9}{1600}\left(\frac{\nu_Q^2}{\nu_L}\right)$\\
\hline
\end{tabular}
\label{table:NQR_freq}
\end{table}
\\When a single rf-frequency (for $\beta$-NMR) or a correlated set of frequencies (for $\beta$-NQR)
are applied such that they correspond to the frequency difference between two adjacent \textit{m}-states,
the polarization is destroyed or reversed. This can be observed as a change in the $\beta$-decay asymmetry.
The latter quantity is defined as
\begin{equation}
\label{asymm}
\mbox{Asymmetry} = \frac{N_{up}-N_{down}}{N_{up}+N_{down}} \simeq \frac{v}{c}A_{\beta}P
\end{equation}
$N_{up}$ is the number of coincident counts detected by the upper
set of detectors, while $N_{down}$ is the number of coincident
counts recorded in the lower scintillators. $A_{\beta}$ is
the asymmetry parameter and depends on the $\beta$-decay
properties of the nucleus under consideration. $P$ is the
initial polarization of the implanted ensemble, induced by the nuclear
reaction.\\
Two techniques are commonly used: the continuous rf technique (CRF) \cite{Arn87,Bor05} and the adiabatic fast passage method (AFP) \cite{Uen05,Kam07}. Both are based on the principle of $\beta$-NMR \cite{Abr61}. In the CRF technique, the spin polarization is resonantly destroyed by mixing the \textit{m} quantum states with a rf-field while the AFP method requires a pulsed production and rf system in order to reverse the population of the \textit{m}-states. This letter describes and compares both techniques which have been applied in the same experimental conditions for the first time.\\
In the CRF method, $^{31}$Al nuclei are continuously implanted in the stopper crystal which is exposed to one ($\beta$-NMR) or several correlated ($\beta$-NQR) rf-frequencies. At the same time, the asymmetry of the $\beta$-decay is measured. When the applied frequency (set) covers the
Larmor frequency ($\beta$-NMR) or the quadrupole coupling constant ($\beta$-NQR), all \textit{m}-states are equally populated. The polarization is destroyed and a resonance is observed when the $\beta$-asymmetry is plotted as a function of the rf-frequency. The resonance amplitude is proportional to the initial polarization \textit{P} of the implanted ensemble.
\\A broad rf-region is scanned in
discrete steps. In order to cover a large frequency range in one step, frequency modulation is used. In case of NQR and for each value of $\nu_Q$, all transition frequencies $\nu_i$ (Table \ref{table:NQR_freq}) are applied simultaneously and
modulated over 55\% of the frequency step between two subsequent
values. Typically, one set of frequencies is applied during several seconds (up to minutes). After going through all
frequency steps and before starting the next scan, data without
rf are taken as a reference. The process is repeated for 30 minutes up to several hours until sufficient statistics are collected.
\\The AFP method uses a pulsed beam which is implanted in the crystal during the time `beam on' as outlined in Fig. \ref{time_scheme}. When the beam is switched off, a single rf-frequency ($\beta$-NMR) or a set of correlated frequencies ($\beta$-NQR) is applied for a few ms (rf-time) and swept over a certain rf-range around a central value. If the sweep range, the rf-field strength and the rf-time are chosen properly, the AFP-condition is fulfilled and the population of the \textit{m}-states is reversed \cite{Abr61}. After the rf pulse sequence, the $\beta$-counting is started, typically for one lifetime. The AFP-cycle is concluded with the same rf-sequence in order to restore the original direction of the polarization. The different NQR frequencies are not applied simultaneously as in the CRF-method but according to a sequence (Fig. \ref{time_scheme}).\\
The duty cycle of the AFP-method is in most cases slightly below 50\%, compared to 100\% in the CRF technique. The resonance amplitude however is proportional to 2\textit{P} (from +\textit{P} to -\textit{P}) when the AFP-condition is exactly met. Because the sensitivity of the method is proportional to the asymmetry change squared, the AFP-method can be up to two times more efficient than the CRF technique.
\\In order to prove that the measured quadrupole coupling constants are independent of the experimental conditions, both methods were applied using a different magnetic field and a different crystal orientation. The continuous rf NMR and NQR were performed in an external field of 0.25 T. The measuring time per frequency was 10 s. For the NQR-measurement, the \textit{c}-axis of the Al$_2$O$_3$ crystal was oriented parallel to the direction of $B_0$. The corresponding transition frequencies are given in the middle column of Table \ref{table:NQR_freq}. All AFP measurements were performed with $B_0$=0.5 T and $\theta$ was put to 90$^\circ$. The transition frequencies are shown in the last column of Table \ref{table:NQR_freq}.

\section{Results}
\label{res} 
\begin{figure}[htb]
\begin{center}
\includegraphics[scale = 0.43]{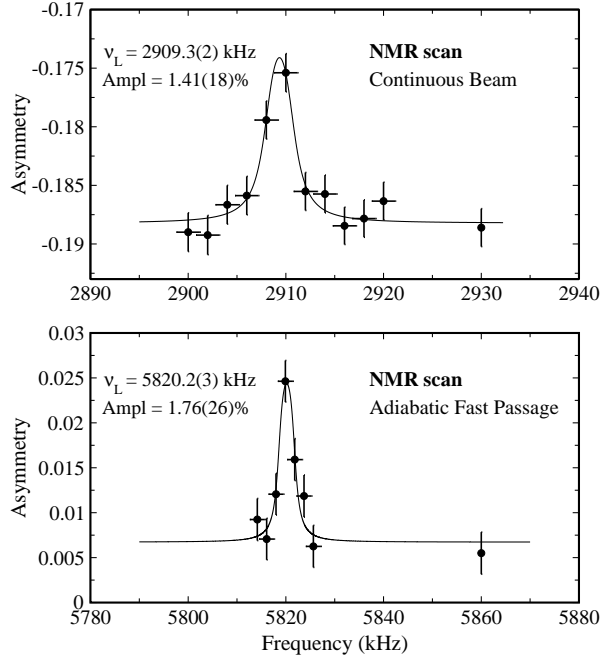}
\caption{$\beta$-NMR resonances obtained with the continuous rf technique (upper panel) and the AFP method (lower panel).}
\label{result_NMR}
\end{center}
\end{figure}

\begin{figure}[htb]
\begin{center}
\includegraphics[scale = 0.46]{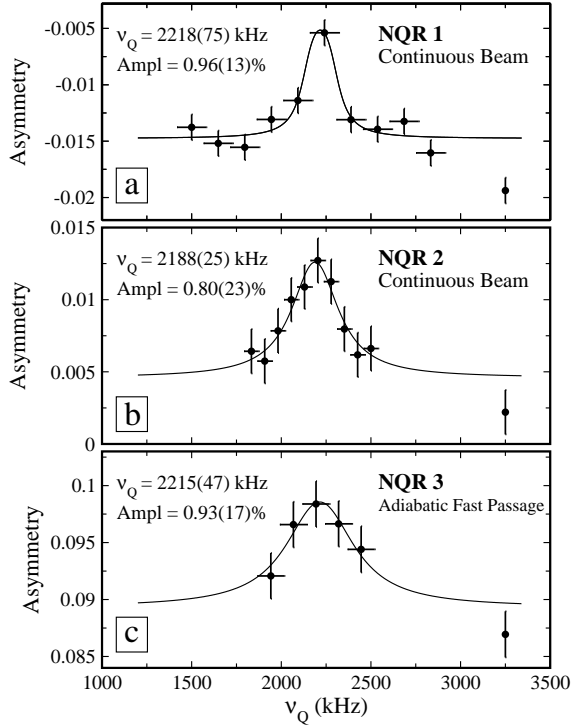}
\caption{$\beta$-NQR resonances obtained with the continuous rf technique (a and b) and the AFP method (c).}
\label{result_NQR}
\end{center}
\end{figure}

In order to implement a successful $\beta$-NQR
experiment on the $^{31}$Al ground state, a precise knowledge of
the Larmor frequency $\nu_{L}$ is required. Therefore, a
$\beta$-NMR measurement (Al implanted in Si) was performed prior to
each NQR-measurement. The results are shown in Fig.
\ref{result_NMR}. The upper panel shows the continuous rf NMR
with a frequency
modulation of 1.2 kHz. Using a fitting function that includes the
Lorentzian line shape and the frequency modulation resulted in $\nu_L$ = 2909.3(2) kHz. Other fit parameters were
the position of the baseline, the FWHM of the curve and the amplitude
(proportional to the polarization).
\\The lower panel of Fig. \ref{result_NMR} displays the AFP result, obtained with 1.54 kHz frequency modulation. The same fitting procedure was applied and a Larmor frequency $\nu_L$ = 5820.2(3) kHz was found.\\
The slightly better relative accuracy obtained with the AFP-method, stems from the fact that the relative
frequency modulation was a little smaller compared to the one applied in the continuous rf NMR, resulting in a narrower resonance and a more
precise value of $\nu_L$. An absolute field calibration was not performed in this
experiment as the NQR transition frequencies $\nu_i$ (see Table
\ref{table:NQR_freq}) do not directly depend on the magnetic
field but only on the Larmor frequency and the unknown quadrupole
coupling constant. Consequently, no magnetic moments were deduced.
\\The spectroscopic quadrupole moment of the $^{31}$Al ground state was initially measured with the continuous rf method.
Starting from a broad scan, covering a wide range of $|$\textit{Q}$|$ and
using a large frequency modulation (81.5 kHz), a resonance was
found at $\nu_Q$=2218(75) kHz (Fig.
\ref{result_NQR}a). In the subsequent NQR, a zoom of the
frequency region was made around the earlier observed resonance
(Fig. \ref{result_NQR}b), applying a frequency modulation of 40.7 kHz. $\nu_Q$=2188(25) kHz was obtained. The continuous rf results were confirmed by a
measurement using the adiabatic fast passage technique with a modulation of 75.5 kHz, resulting in $\nu_Q$=2215(47) kHz (Fig.
\ref{result_NQR}c). The weighted mean of the three $\nu_Q$'s is calculated to be
$\overline{\nu}_Q\left(^{31}\mbox{Al}\right)$=2196(21) kHz. Since the quadrupole
coupling constant of stable $^{27}$Al was also determined in an
$\alpha$-Al$_2$O$_3$ crystal at room temperature, its value can be directly compared to $\overline{\nu}_\textit{Q}\left(^{31}\mbox{Al}\right)$. Using the known values $|$\textit{Q}($^{27}$Al)$|$ = 146.6(10) mb \cite{Kel99} and $\nu_{Q}$($^{27}$Al) = 2402.5(17) kHz \cite{Fil97}, the quadrupole
moment of the $^{31}$Al ground state can be calculated as follows:
\begin{equation}
|Q\left(^{31}\mbox{Al}\right)| = \frac{|Q\left(^{27}\mbox{Al}\right)|\;\overline{\nu}_Q\left(^{31}\mbox{Al}\right)}{\nu_Q\left(^{27}\mbox{Al}\right)}
\label{q_moment_31Al}
\end{equation}
$|Q(^{31}$Al)$|$ =
134.0(16) mb is obtained. This result is more than
one order of magnitude more precise and in agreement with the earlier measured quadrupole moment of $^{31}$Al (112(32) mb \cite{Nag08}).
\\A comparison of the NMR amplitudes given in Fig. \ref{result_NMR} shows no large difference between the AFP-resonance and the continuous rf result. Within the errors, both amplitudes are the same. Also for the NQR measurements (indicated in Fig. \ref{result_NQR}) no clear difference in amplitude exists. The expected double effect has not been observed since the AFP-technique was not optimized and, as a consequence, the exact AFP-condition was never reached.

\section{Discussion}
\label{disc}
\begin{figure}[htb]
\begin{center}
\includegraphics[scale = 0.31]{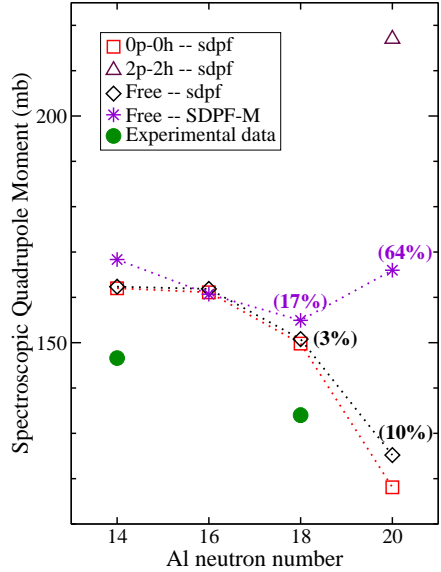}
\caption{(Color online) An overview of the calculated and the measured quadrupole moments of the neutron-rich odd-mass Al ground-states (\textit{I}=5/2$^+$). All calculations were made using \textbf{the effective charges $\mathbf{e_p = 1.3 e}$ and $\mathbf{e_n = 0.5 e}$}. The contribution of intruders configurations in the ground state is given between brackets (if different from zero). }
\label{Al_calc}
\end{center}
\end{figure}

\begin{figure}[htb]
\begin{center}
\includegraphics[scale = 0.31]{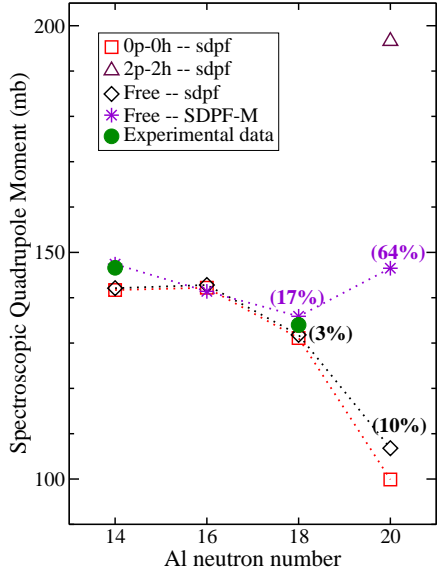}
\caption{(Color online) An overview of the calculated and the measured quadrupole moments of the neutron-rich odd-mass Al ground-states (I=5/2$^+$). All calculations were made using \textbf{the effective charges $\mathbf{1.1 e}$ and $\mathbf{0.5 e}$} for protons and neutrons respectively.}
\label{Al_calc_1.1_0.5}
\end{center}
\end{figure}

Recently, the proton core polarization charges and the electric quadrupole moments of the neutron-rich Al isotopes have been investigated in the microscopic particle-vibration coupling (PVC) model \cite{Yos09}. The single particle wave functions used in this model are obtained from solving the Skyrme Hartee-Fock-Bogoliubov equation self-consistently while the phonon energies are deduced from the quasiparticle-random-phase approximation. The $^{31}$Al quadrupole moment was calculated for two different values of the Landau-Migdal (LM) interaction strength, resulting in $Q_{PVC}(^{31}$Al) = 136 mb and 139 mb, in excellent agreement with our experimental value. By comparing the calculated single particle quadrupole moment (without coupling to E2-excitations) with the quadrupole moment calculated in the PVC model, the authors deduced a proton core polarization charge of $1.03 e$ and $1.08 e$ for $^{31}$Al, depending on the LM interaction strength used.\\
In this letter, the $^{31}$Al quadrupole moment is interpreted in the framework of two large-scale shell-model approaches. ANTOINE calculations \cite{Cau99}
are performed with the sdpf residual interaction \cite{Num01} and the Monte Carlo Shell Model (MCSM) \cite{Hon95,Miz96} is used with the
SDPF-M interaction \cite{Uts04}. Both models are very efficient in outlining the nuclear properties of isotopes that belong to the island
of inversion. For the odd-mass neutron-rich Al-isotopes, four different calculations are made and presented in Fig. \ref{Al_calc}. Three of them are ANTOINE calculations: one with all neutrons confined to the sd-shell (0p-0h, squares), one with two neutrons forced into the f$_{7/2}$p$_{3/2}$ orbitals (2p-2h, triangles) and one without truncations in the $\nu$(sdf$_{7/2}$p$_{3/2}$) model space (free sdpf, diamonds). The fourth dataset is a MCSM calculation using the SDPF-M interaction in the $\nu$(sdf$_{7/2}$p$_{3/2}$) valence space (free SDPF-M, stars). In the ANTOINE calculations, protons are restricted to the sd-shell while the MCSM imposes no restrictions on the proton space. All calculations are performed using the standard effective charges $e_p = 1.3 e$ and $e_n = 0.5 e$ \cite{Bro88}. The experimental data (dots) are taken from this work and from Ref. \cite{Kel99}.\\
\\For $^{27-31}$Al, 0p-0h calculations and calculations in the untruncated model space predict similar quadrupole moments. No intruder configurations are present in the ground states of $^{27-29}$Al while a minor contribution (see Fig. \ref{Al_calc}) is predicted for $^{31}$Al. The theoretical Q-moments reproduce the observed trend but theory overestimates all experimental values by about 15\%.\\
As the quadrupole moments of the odd-mass Al-isotopes are dominated by the odd d$_{5/2}$ proton hole, the proton effective charge has been rescaled to $e_p = 1.1 e$ in order to obtain a quantitative agreement with the experimental value of $^{27}$Al. Fig. \ref{Al_calc_1.1_0.5} gives an
overview of the experimental and rescaled theoretical Q-moments of $^{27-33}$Al, adopting the same color code as in Fig. \ref{Al_calc}.
\\All theoretical predictions for $^{31}$Al (\textit{N}=18), calculated with the reduced effective proton charge, show a good agreement with the experimental value. Pure $\nu$ sd-calculations as well as calculations in the untruncated model space can explain the observed Q-moment. This indicates that the $^{31}$Al ground state is dominated by normal sd-configurations with a possible small admixture of intruder states.\\
For $^{33}$Al (\textit{N}=20), free ANTOINE and MCSM calculations predict an amount of 10\% and 64\% respectively of intruder configurations in the ground state, leading to an enhanced quadrupole moment compared to what is suggested by 0p-0h calculations. In case $^{33}$Al has a pure sd ground-state configuration, a strong decrease of the quadrupole moment with respect to the $^{31}$Al value should be observed, as expected for a `normal' closed-shell nucleus. However, a recent report on the g-factor of $^{33}$Al suggests a non-negligible contribution of neutron excitations across \textit{N}=20 in the ground state \cite{Him06}. A measurement of the $^{33}$Al quadrupole moment will therefore provide decisive information about the structure of the Al isotopes at the \textit{N}=20 shell closure.
\begin{figure}[htb]
\begin{center}
\includegraphics[scale = 0.35]{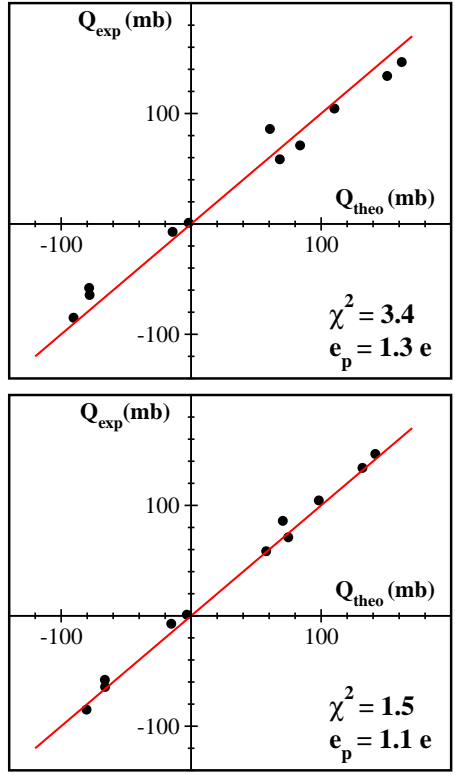}
\caption{Known quadrupole moments of $\pi$sd-shell nuclei with odd Z and even N. The theoretical values are obtained using the sdpf residual interaction with the protons restricted to the sd-shell and the neutrons allowed in the sdf$_{7/2}$p$_{3/2}$ space. Scattering around $Q_{exp}=Q_{theo}$ is studied. In the upper panel, $Q_{theo}$ is calculated with $e_p=1.3e$ while the lower panel shows $Q_{theo}$ values determined with $e_p=1.1e$. The experimental errors are smaller than the symbol size.}
\label{all_sd_1.3_and_1.1_0.5}
\end{center}
\end{figure}
\\The use of the reduced effective proton charge $e_p=1.1 e$ to describe the odd-mass Al quadrupole moments can be justified in the framework of the full $\pi$sd shell. The standard effective charges $e_p=1.3 e$ and $e_n=0.5 e$ were established in 1988 by Brown and Wildenthal \cite{Bro88} based on a comparison of shell-model calculations with the experimental E2 $\gamma$-decay transition matrix elements and the not very precise static quadrupole moments known at that time. Since then, nine new and/or more precise static quadrupole moments of odd-Z even-N $\pi$sd-shell nuclei became available, including the present result for $^{31}$Al. Isotopes with an odd proton and an even neutron number are excellent probes to define the proton effective charge used in the $\pi$sd-shell as their quadrupole moments are mainly determined by proton configurations.\\
In Fig. \ref{all_sd_1.3_and_1.1_0.5}, each odd-Z even-N $\pi$sd-isotope for which the ground-state quadrupole moment is measured, is represented by a data point with $Q_{theo}$ as X-coordinate and $Q_{exp}$ as Y-coordinate. All data scatter around the $Q_{exp}=Q_{theo}$ line. In the upper panel, the quadrupole moments are calculated using the ANTOINE sdpf residual interaction with protons in the sd shell and neutrons in the sdf$_{7/2}$p$_{3/2}$ orbits using the standard effective charges $e_p=1.3e$ and $e_n=0.5 e$. A rather large reduced $\chi^2$-value of 3.4 is found with respect to the $Q_{exp}=Q_{theo}$ curve. In the lower panel, $Q_{theo}$ is determined using the reduced effective proton charge $e_p=1.1e$. A much better agreement between experiment and theory ($\chi_{red}^2 = 1.5$) is observed. The experimental data are taken from this work and from Refs. \cite{Sto05, Mat05}.\\
Thus, based on a compilation of recent and remeasured quadrupole moments for odd-Z even-N $\pi$sd isotopes, it can be concluded that $e_p = 1.1e$ is a more realistic effective proton charge for the sd-shell than the previously adopted value $1.3e$. This conclusion offers a strong argument to justify the use of the effective proton charge $1.1e$ in the discussion of the $^{31}$Al quadrupole moment.

\section{Summary}
\label{summ}
In conclusion, the electric quadrupole coupling constant of $^{31}$Al, produced in a projectile-fragmentation reaction, was measured using two
nuclear quadrupole resonance techniques, one based on a continuous rf-signal, the other on a pulsed sequence. The measured values are in good agreement with each other, leading to a mean value of $\nu_{Q}$=2196(21) kHz. From this, the quadrupole moment of $^{31}$Al could be extracted: $\left|Q(^{31}\mbox{Al})\right|$ = 134.0(16) mb.\\
The precise value of $|Q(^{31}$Al)$|$ together with other experimental quadrupole moments of odd-Z even-N sd-nuclei were used to determine the proton effective charge in the $\pi$sd-shell. Calculations performed with $e_p = 1.1e$ result in a better agreement with experiment than calculations using the standard effective proton charge $e_p = 1.3e$.\\
Comparison with shell-model calculations using the sdpf and SDPF-M residual
interactions shows that the ground state of $^{31}$Al is dominated by normal sd-shell configurations with a possible small admixture
of neutron 2p-2h states. This points to a gradual transition from the deformed Mg isotopes to the normal Si ground states. A measurement of $Q(^{33}$Al) would add important information to the study of the nuclear structure of the odd-mass Al-isotopes at the border
of the island of inversion. The accurate value of $|Q(^{31}$Al)$|$, obtained in this work, will contribute to a correct interpretation of that new result.\\\\
We are grateful to the GANIL staff for the technical support and to Dr. E. Yagi for the useful help and
advice with the X-ray diffraction analysis of the $\alpha$-Al$_2$O$_3$
sample. This work
has been financed by the European Community FP6 - Structuring the
ERA - Integrated Infrastructure Initiative contract EURONS No.
RII3-CT-2004-506065, by the
FWO-Vlaanderen and by the IAP-programme of the Belgium Science Policy
under grand number P6/23.

\end{document}